\documentclass[fleqn,usenatbib]{mnras}

\usepackage{newtxtext,newtxmath}
\usepackage[T1]{fontenc}
\usepackage{ae,aecompl}

\usepackage{graphicx}	% Including figure files
\usepackage{amsmath}	% Advanced maths commands
\usepackage{amssymb}	% Extra maths symbols

\title[Scattering variability of the Crab pulsar]{Scattering features and variability of the Crab pulsar}

\author[L.\,N.\,Driessen et al.]{
L.\,N.\,Driessen$^{1,2,3}$\thanks{Contact e-mail: Laura@Driessen.net.au (LND)},
G.\,H.\,Janssen$^{1,4}$,
C.\,G.\,Bassa$^{1}$,
B.\,W.\,Stappers$^{3}$,
and D.\,R.\,Stinebring$^{5}$
\\
$^{1}$ASTRON, the Netherlands Institute for Radio Astronomy, Oude Hoogeveensedijk 4, 7991 PD, Dwingeloo, The Netherlands\\
$^{2}$Anton Pannekoek Institute for Astronomy, University of Amsterdam, Science Park 904, 1098 XH Amsterdam, The Netherlands\\
$^{3}$Jodrell Bank Centre for Astrophysics, School of Physics and Astronomy, The University of Manchester, Manchester, M13 9PL, UK\\
$^{4}$Department of Astrophysics/IMAPP, Radboud University, P.O. Box 9010, 6500 GL Nijmegen, The Netherlands\\
$^{5}$Department of Physics and Astronomy, Oberlin College, 110 No. Professor St., Oberlin OH 44074, United States
}

\date{Accepted 2018 November 17. Received 2018 November 09; in original form 2018 April 09}

\pubyear{2018}

\begin{document}
\label{firstpage}
\pagerange{\pageref{firstpage}--\pageref{lastpage}}
\maketitle

% Abstract of the paper
\begin{abstract}
We report on Westerbork Synthesis Radio Telescope observations of the Crab pulsar at $350\,\mathrm{MHz}$ from 2012 November 24 until 2015 June 21. 
During this period we consistently observe variations in the pulse profile of the Crab. Both variations in the scattering width of the pulse profile as well as delayed copies, also known as echoes, are seen regularly. These observations support the classification of two types of echoes: those that follow the truncated exponential shape expected for the thin-screen scattering approximation, and echoes that show a smoother, more Gaussian shape. During a sequence of high-cadence observations in 2015, we find that these non-exponential echoes evolve in time by approaching the main pulse and interpulse in phase, overlapping the main pulse and interpulse, and later receding.
We find a pulse scatter-broadening time scale, $\tau$, scaling with frequency as $\nu^{\alpha}$, with $\alpha=-3.9\pm0.5$, which is consistent with expected values for thin-screen scattering models.
\end{abstract}


%%%%%%%%%%%%%%%%%%%%%%%%%%%%%%%%%%%%%%%%%%%%%%%%%%

%%%%%%%%%%%%%%%%% BODY OF PAPER %%%%%%%%%%%%%%%%%%

\section{Introduction}
\label{sec:Intro}

\begin{figure*}
	\includegraphics[width=0.99\textwidth]{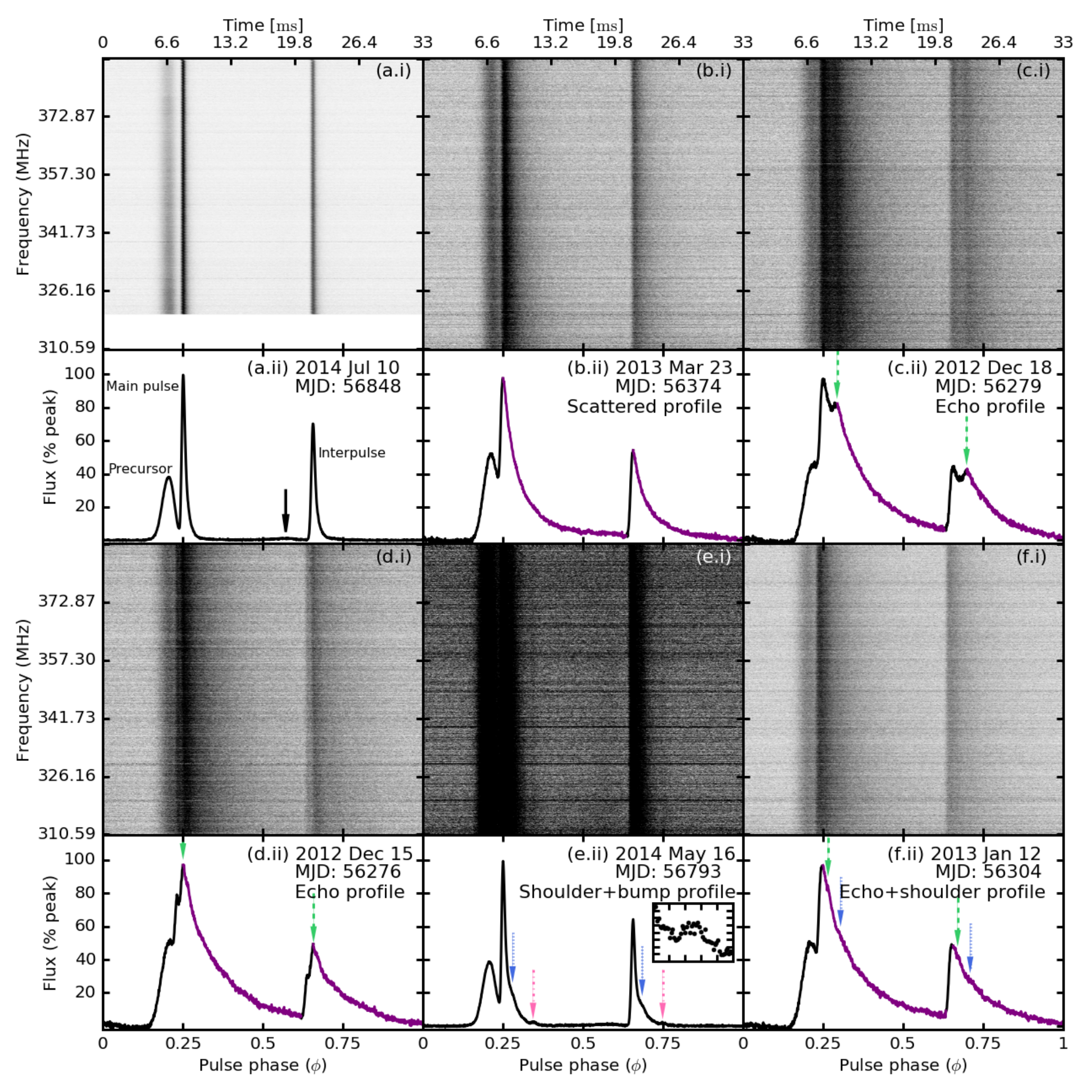}
    \caption{WSRT observations of the Crab pulsar at $350 \, \mathrm{MHz}$. Each grey scale plot, showing intensity as a function of frequency and pulse phase, corresponds to the frequency and time averaged pulse profile below. Green (dashed) arrows indicate echo features (c, d and f), blue (dotted) arrows indicate shoulder features (e and f), and  pink (dot-dashed) arrows indicate bump features (e). In the grey scale plots the grey scale indicates how bright the features are, with all of the observations scaled linearly such that the maximum is 1.0 (black) and the minimum is 0.0 (white). In plot (e) the grey scale has been adjusted to a minimum of 0.0 and maximum of 0.2 in order to show the shoulder and bump features clearly. In (a) the intrinsic features of the pulse profile of the Crab pulsar at $350\,\mathrm{MHz}$ are labelled, including the low-frequency precursor to the IP \citep{mh96, ksl12} which is indicated with a solid black arrow. (b) shows an example of an observation with additional pulse scatter-broadening (purple). (c) shows an example of an observation with a clear echo feature. In (d) the echo is brighter than the main profile. Plot (e) shows observations with (blue/dashed) shoulder and (pink/dot-dashed) bump features, the inset figure in (e.ii) shows a zoom in on the bump trailing the MP. Plot (f) shows an observation with significant pulse scatter-broadening, an echo (close in phase to the main profile), and a shoulder.}
    \label{fig: figure 1}
\end{figure*}

Pulsar B0531+21 is located within the Crab nebula (M1) and is commonly referred to as the Crab pulsar. The Crab pulsar was formed in a core-collapse supernova which was observed by astronomers in 1054\,AD. The pulsar itself was discovered in 1968 \citep{1968Sci...162.1481S}. The Crab pulsar is visible at all observable wavelengths and is one of the brightest objects in the radio sky \citep[e.g.][]{2010A&A...515A..36K}. It has a rotation period of $33.7\,\mathrm{ms}$ and is $2\,\mathrm{kpc}$ away \citep{2008ApJ...677.1201K}. The pulse profile of the Crab pulsar has three main components, which are shown in Fig.\,\ref{fig: figure 1}a: the main pulse (MP) with a precursor, and the interpulse (IP) at a $\sim145^{\circ}$ phase offset from the peak of the MP. The Crab pulse profile is known to vary intrinsically over frequency \citep[e.g.][and references therein]{mh96}. 

The Crab pulsar is known to show variability in pulse scatter-broadening, which is seen as broadening of the trailing parts of the profile. Apart from general variability in pulse scatter-broadening, in 1974 \citep{1975MNRAS.172...97L}, 1997 \citep{2000ApJ...543..740B, 2001MNRAS.321...67L}, and 2006 \citep{2008A&A...483...13K}, anomalous increases in pulse scatter-broadening of the Crab pulsar were observed.
During the 1974 and 1997 periods of additional pulse scatter-broadening, a new feature was observed as an extra peak
on the trailing edge of both the MP and IP. In these papers, extra features were labelled as ``ghosts" or ``echoes". 
An echo was not seen on the precursor as it has a much lower flux density compared to the MP
and any extra feature would blend into the MP \citep{2000ApJ...543..740B}.
During these periods of increased pulse scatter-broadening an echo would appear as a delayed copy of the pulse profile and over time the delay would decrease until the echo and main profile overlap, after which the delay would increase again.
It was observed in 1974 and 1997 that while the echo persisted, the peak flux density
of the pulse profile decreased, likely due to the effect of pulse scatter-broadening in combination with an increase in dispersion measure \citep{GS2011}.

Generally the pulse scatter-broadening that is seen on the Crab is attributed to the interstellar medium (ISM), while the echo feature is attributed to a second scattering screen \citep[e.g.][]{2000ApJ...543..740B} or lens \citep{GS2011} close to the Crab pulsar. The Crab pulsar is surrounded by the Crab nebula, which is a plerionic supernova remnant consisting of a pulsar wind nebula surrounded by cold ejecta dust, without the forward and reverse shock typical of young supernova remnants \citep{2015ApJ...801..141O}. It is therefore logical to attribute this extra scattering screen
to material in the nebula \citep[e.g.][]{2008A&A...483...13K}.

\begin{figure}
	\includegraphics[width=0.49\textwidth]{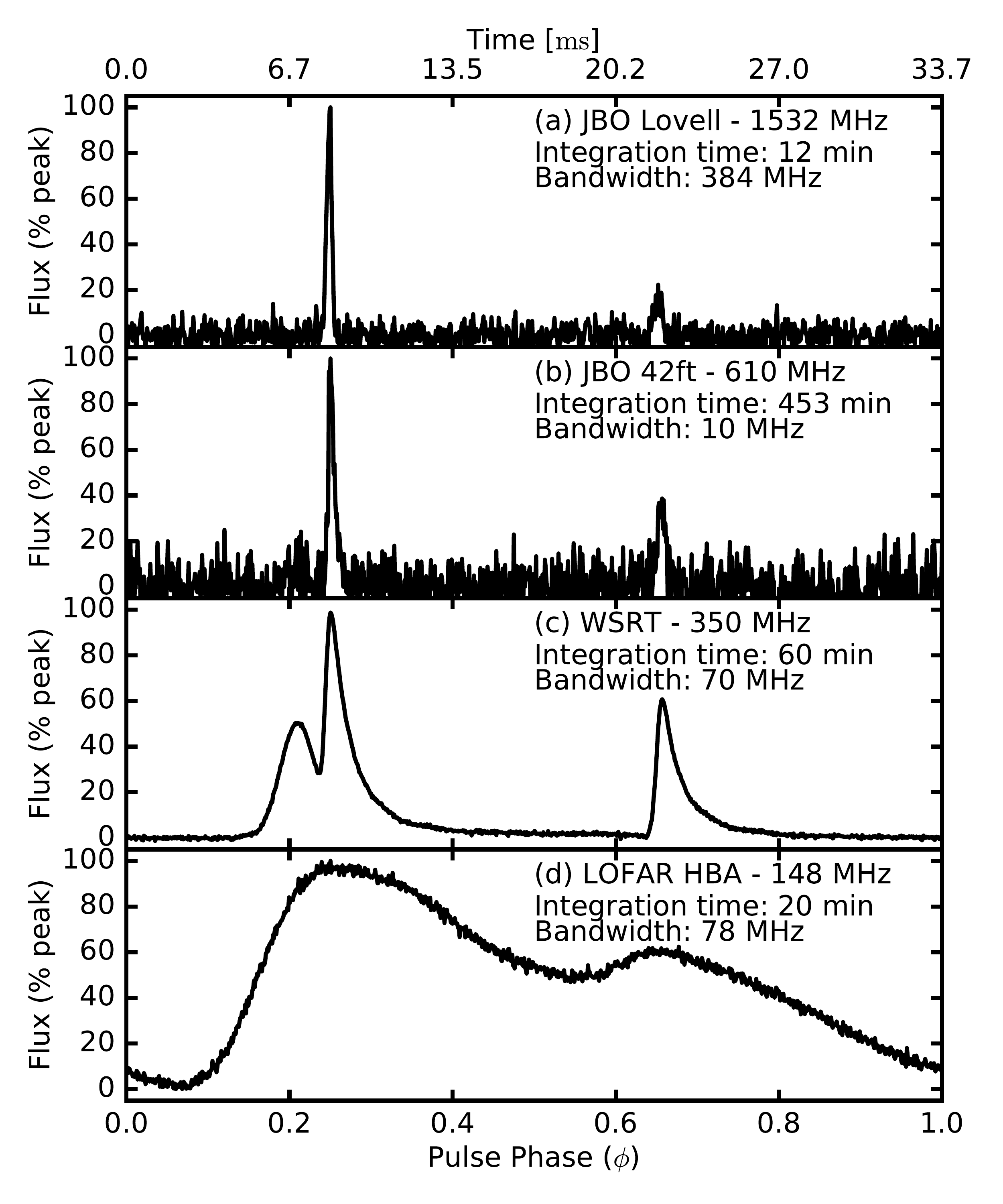}
    \caption{
    Observations of the Crab pulsar from (a) the Lovell telescope at the Jodrell Bank Observatory 
    (JBO), (b) the 42ft telescope at JBO, (c) the WSRT, and
    (d) the LOw Frequency ARray (LOFAR) High Band Antennas (HBA).
    The Lovell observation was taken on 2015 June 10,
    the 42ft and WSRT observations were both taken on 2015 June 11,
    the LOFAR observation was taken on 2015 June 12. The effect of pulse scatter-broadening increases as the 
    observing frequency decreases. Apart from the effects of pulse scatter-broadening, the profile evolution with 
    frequency can be seen with the changes in the precursor to the MP.
    For illustration purposes, the profiles have been scaled such that the peak of the MP is
    defined at 100\%, and the lowest part of the profile is at zero. The intrinsic flux
    densities are different for each observing frequency. Note that for both the WSRT
    and LOFAR profiles the baseline is increased due to the pulse scatter-broadening time scale covering
    significant fractions of the pulse period. The profiles have also been aligned such that
    the MP is at a pulse phase of 0.25.
    }
    \label{fig: figure 2}
\end{figure}

Pulse scatter-broadening is a multi-path propagation effect caused by inhomogeneity in electron density between a pulsar
and the observer \citep[e.g.][]{1969Natur.221..158R,1970ApJ...162..707R,1986ApJ...311..183C}. The change in path length of light due
to material with irregular densities causes a delay in the arrival time of the light
from the pulsar. 
The effect of scattering on pulse profiles is more pronounced at lower
frequencies as pulse scatter-broadening time, $\tau$, is related to frequency, $\nu$, by
$\tau \propto \nu^{\alpha}$. 
There are several models that are generally used to describe the effects of scattering on a pulsed signal. A thin-screen scattering model is an approximation that assumes that the material causing the pulse scatter-broadening is in a thin screen between the observer and the object \citep{1972MNRAS.157...55W}. 
The screen is usually attributed to the ISM and it is expected that $-4.4<\alpha<-4$ \citep{2015MNRAS.454.2517L}; where $\alpha=-4.4$ assumes a Kolmogorov spectrum \citep{2002AstL...28..251K} and $\alpha=-4$ assumes Gaussian inhomogeneities \citep{1971ApJ...164..249L, 1976ApJ...206..735L}, however many pulsars have been observed to have $\alpha$ values greater than $-4$ and less than $-4.4$ \citep{2017ApJ...846..104K}. 
The frequency dependence of pulse scatter-broadening is shown in Fig.\,\ref{fig: figure 2}. 
In a pulse profile, the effect of thin-screen scattering appears as a truncated exponential or ``exponential tail", which is a broadening of the trailing edge (or ``tail") of the profile components \citep[e.g.][]{1975MNRAS.172...97L,2000ApJ...543..740B}.

In this paper we present and analyse additional pulse scatter-broadening and variable features
in the pulse profile of the Crab pulsar in Westerbork Synthesis Radio Telescope \citep[WSRT;][]{1974A&A....31..323B} observations at $350\,\mathrm{MHz}$. In Sec.\,\ref{sec:Obs} we describe our observations and our method for modelling the pulse profiles. In Sec\,\ref{sec:results} we present our results and in Sec.\,\ref{sec:Discussion} we discuss the features we see in the pulse profiles. We conclude in Sec.\,\ref{sec:Conclusion}.

\section{Observations and analysis}
\label{sec:Obs}

\subsection{Observations}
We have obtained observations of the Crab pulsar with the WSRT  at
observing frequencies between 310 and 380\,MHz with the
\texttt{PuMa\,II} pulsar instrument \citep{2008PASP..120..191K}. \texttt{PuMa\,II} 
records real valued, Nyquist sampled timeseries for dual
polarizations for 8 subbands of 10\,MHz bandwidth each. The subbands
were overlapped by 1.25\,MHz to mitigate a roll-off of the bandpass,
resulting in a total effective bandwidth of 70\,MHz. The Nyquist
sampled timeseries were coherently dedispersed to a DM\footnote{Dispersion measure (DM) is a measure of the column density of free electrons between
 a pulsar and the observer. DM causes a delay in the pulse arrival time, which
 is more pronounced at lower frequencies. The time delay is related to frequency
 by $t \propto \nu^{-2}$.} of
56.79\,pc\,cm$^{-3}$ and folded with the
\texttt{dspsr}\footnote{\href{http://dspsr.sourceforge.net/}{http://dspsr.sourceforge.net}}
\citep{2011PASA...28....1V} software package, resulting in 156.25\,kHz wide
channels and 1024 pulse phase bins across the profile.  A total of 89
observations with integration times varying from 20 to 60\,min were
taken between 2012 November 24 and 2015 June 21. We excluded 5
observations that were affected by high sky temperatures due to the
Sun passing near the line of sight towards the Crab pulsar (each year
around June 15). Due to the interferometric nature of the WSRT array,
very little radio frequency interference (RFI) was present in the
folded data. Remaining narrow-band RFI was flagged manually using
tools from the
\texttt{psrchive}\footnote{\href{http://psrchive.sourceforge.net/}{http://psrchive.sourceforge.net}}
software suite \citep{2004PASA...21..302H}. As the DM towards
the Crab pulsar varies with time \citep[e.g.][]{1973ApJ...181..875R,2008A&A...483...13K}, we
corrected the DM of each observation to the value that minimized the 
width of the profile components.

\subsection{Pulse profile features}
The observations of the Crab pulsar show significant variations in the
pulse scatter-broadening properties in the $350\,\mathrm{MHz}$ observing band. A set of representative observations are shown
in Fig.\,\ref{fig: figure 1} as time averaged and frequency collapsed profiles,
and as profiles across a range of frequencies. The time averaged and frequency collapsed profiles for all 84 observations are shown in Appendix\,\ref{App: A}. In observations with low
pulse scatter-broadening, such as those in Fig.\,\ref{fig: figure 1}a and
Fig.\,\ref{fig: figure 1}e, the precursor to the MP is almost completely separated
from the MP, while in observations with high pulse scatter-broadening the
two components merge. Low pulse scatter-broadening observations also show the
presence of the low-frequency precursor to the IP (indicated by a black arrow in Fig.\,\ref{fig: figure 1}a, which has previously been seen at
observing frequencies predominantly below 200\,MHz 
\citep[e.g.][and references therein]{ksl12} and weakly at 330\,MHz
\citep{mh96}.

\begin{figure*}
	\includegraphics[width=0.99\textwidth]{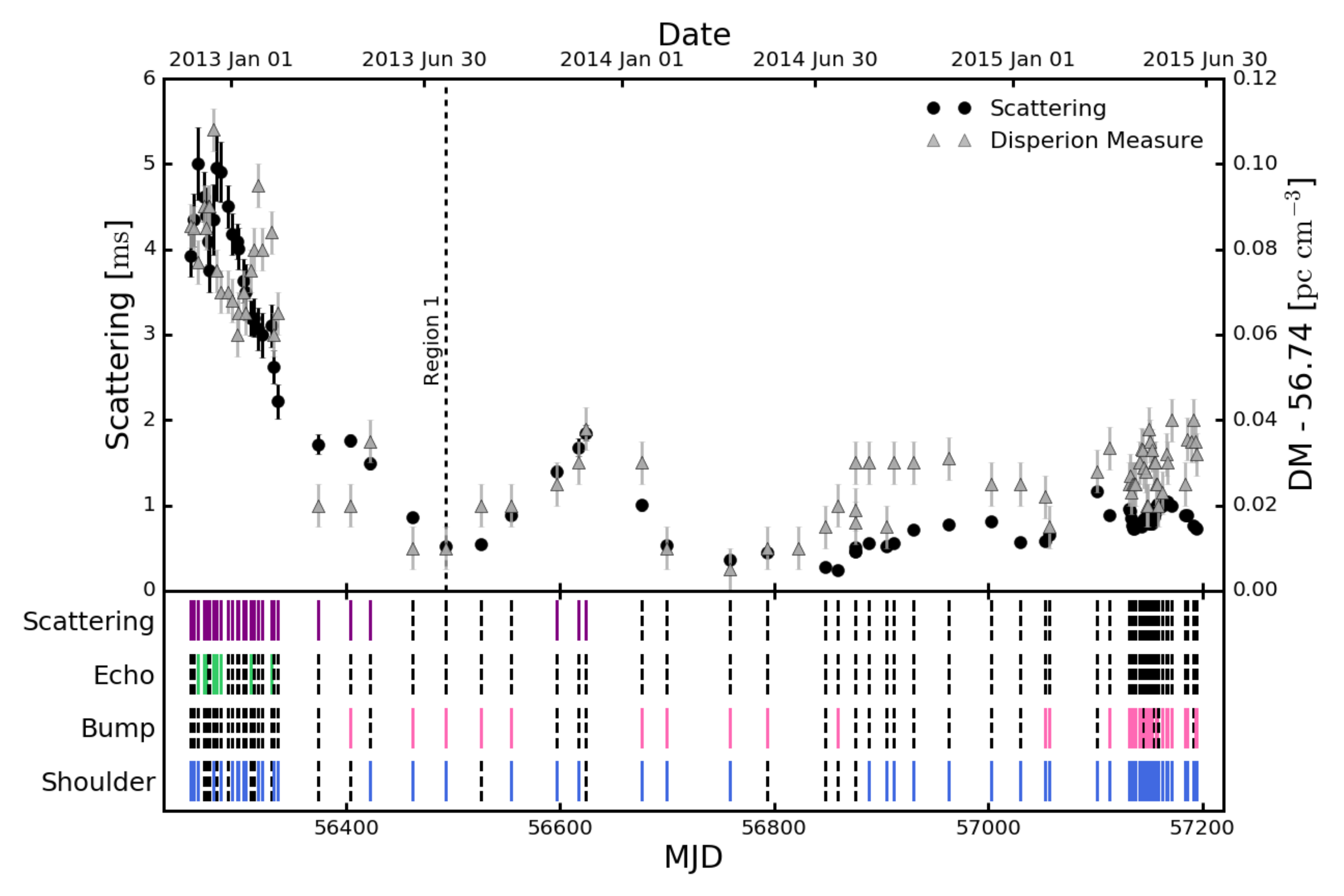}
    \caption{Properties of the Crab pulsar at $350 \, \mathrm{MHz}$ over time. The top panel shows the DM and pulse scatter-broadening, using a single-screen model. The bottom panel shows how often certain features occur. In the bottom panel a black dashed mark means that the feature is not present in that observation, a solid coloured mark means that the feature is present. Examples of these features are shown in Fig.\,\ref{fig: figure 1} (with the same colour coding). Some features, particularly bumps and shoulders, may be hidden during periods of increased pulse scatter-broadening.}
    \label{fig: figure 3}
\end{figure*}

Several observations show delayed copies, or echoes, of the pulse
profile on the trailing edge of both the MP and IP. These copies are
complete copies of the pulse profile; there is always an echo on the
IP if there is one on the MP and vice versa, and there can be one
echo, multiple echoes, or no echoes in an observation. Some echoes,
such as the echo shown in Fig.\,\ref{fig: figure 1}c, have a sharp
rising edge and scatter-broadened trailing edge; we will call these
exponential echoes. Other echoes, such as those in
Fig.\,\ref{fig: figure 1}e, have small amplitudes, lack a sharp
rising edge, and have a more Gaussian-like shape; we will call
these non-exponential echoes.  For exponential echoes the echo of the
precursor is not seen as it overlaps the MP and for non-exponential
echoes the amplitude of the precursor echo is not seen as it is below
the noise.  For clarity, we have distinguished between two types of
non-exponential echo, which we will call ``shoulders'' and ``bumps'', examples of which are shown in
Fig.\,\ref{fig: figure 1}e and f. We will use the term ``echo'' to refer to
the exponential echoes. Fig.\,\ref{fig: figure 3} shows the epochs where we observe echoes, shoulders,
and bumps in our WSRT observations.

\subsection{Pulse profile modelling}
\label{sec: pulse profile modelling}
To model the observed pulse profiles and constrain pulse scatter-broadening
properties, we implicitly assume that the pulse scatter-broadening can be modelled
with the thin-screen scattering approximation. Hence, we model the
time-dependent pulse scatter-broadening function as a truncated exponential of the
form $s(t)=\frac{1}{\tau}\exp\bigl(-\frac{t}{\tau}\bigr)H(t)$
\citep[e.g.][]{2014PASP..126..476M}, where $\tau$ is the pulse scatter-broadening
time scale, and $H(t)$ the Heaviside step function as a function of
time $t$. 

The components of the intrinsic pulse profile of the Crab pulsar are
modelled with von Mises functions (circular normal distributions),
$f(\phi)=A\exp\bigl(\kappa[\cos{2\pi(\phi-\phi_0)}]\bigr)$ where $\phi$ is pulse phase, for a
component at pulse phase $\phi_0$, amplitude $A$, and reciprocal width
$\kappa$. The reciprocal width is related to the Gaussian width
$\sigma$ through $\kappa^{-1}\approx\sigma^2$. The intrinsic profile
is modelled as the sum of three von Mises functions, fitting the pulse
phase, amplitude, and reciprocal of the width of the MP, IP, and
precursor.

As the pulse scatter-broadening time scale $\tau$ can exceed the spin period ($P=33.7\,\mathrm{ms}$) of the Crab pulsar (e.g.\ Fig.\,\ref{fig: figure 1}c), we have to
ensure that the scattered flux located outside the pulse phase range
$0<\phi<1$ is taken into account. To achieve this, the intrinsic
profile is defined for one rotation and padded with zeros for several
preceding and following rotations. The number of rotations depends on
the pulse scatter-broadening time scale $\tau_\phi$ (defined in pulse phase as
$\tau_\phi=\tau/P$), such that the pulse scatter-broadening function at large $t$
becomes negligible. The zero padded intrinsic profile was subsequently
convolved with the pulse scatter-broadening function. Finally, all rotations were
summed to obtain the scattered profile \citep{2016MNRAS.462.2587G}.

To determine the intrinsic parameters of the Crab pulsar profile, we
fit scattered model profiles to the four observations with the lowest
pulse scatter-broadening time scales. These observations were
fully averaged in time and collapsed in frequency to 32 frequency
channels of 2.5\,MHz in width. The parameters of the intrinsic profile
of each channel were fitted simultaneously with the pulse scatter-broadening
time scale $\tau_\phi$, and an offset in the baseline flux. The pulse scatter-broadening
time scale in these four observations varied between 0.25 to 0.47\,ms
at a reference frequency of $350\,\mathrm{MHz}$, with
pulse scatter-broadening powerlaw indices of $-3.8$ to $-4.2$. Flux
uncertainties were estimated from the off-pulse region at pulse phases
$0.0<\phi<0.1$ (with the MP at $\phi=0.25$). We find that the
phase offset of the IP and the precursor with respect to the
MP, as well as the widths of the three components, do not
vary significantly with frequency over our available bandwidth.
The relative amplitudes of the IP and
precursor, measured with respect to the MP, do vary with
frequency over our observed bandwidth, decreasing by 0.12\,percent per MHz. The fitted values
are given in Table\,\ref{tab: table 1} and are compared to previously derived values. We expect some difference between our values and those derived at $610\,\mathrm{MHz}$ as the intrinsic profile of the Crab is known to vary with frequency \citep[e.g.][]{GS2011}. Our values match well with those derived by \citet{2000ApJ...543..740B} at $350\,\mathrm{MHz}$, apart from some difference between the precursor position and width. We use our intrinsic profile parameters
to create a frequency dependent model for the intrinsic pulse profile
of the Crab pulsar that is valid for our observing bandwidth. 

\begin{table*}
	\centering
	\caption{The parameters of the intrinsic profile from our model at $350 \,\mathrm{MHz}$ compared to the results of \citet{2000ApJ...543..740B} and \citet{2013Sci...342..598L}. The amplitude values are relative to the MP, which has an amplitude set to $1.0$. The position is the phase position in degrees and is measured relative to the MP position. The errors quoted here are $1\sigma$ errors.}
	\begin{tabular}{lllrrr}
		\hline
		 & Frequency & Component & Amplitude & Position ($\degr$) & Width ($\degr$)\\
        \hline
        Our results & $350\,\mathrm{MHz}$ & Main pulse & 1.0 & $0.0$ & $2.85(11)$\\
         & & Interpulse & 0.623(2) & $145.74(4)$ & $3.36(10)$\\
         & & Precursor & 0.210(1) & $-18.25(11)$ & $13.9(2)$\\
        \hline
        \citet{2000ApJ...543..740B} & $327\,\mathrm{MHz}$ & Main pulse & $1.0(8)$ & $0.00(5)$ & $2.80(1)$\\
         & & Interpulse & $0.67$ & $145.72$ & $3.5(1)$\\
         & & Precursor & $0.29(5)$ & $-17.8(5)$ & $16.4(6)$\\         
         & $610\,\mathrm{MHz}$ & Main pulse & $1.00(8)$ & $0.00(14)$ & $3.3(3)$\\
         & & Interpulse & $0.48$ & $145.6$ & $4.1(3)$\\
         & & Precursor & $0.06(2)$ & $-19.4(9)$ & $12(4)$\\
        \hline
		\citet{2013Sci...342..598L} & $610\,\mathrm{MHz}$ & Main pulse & 1.0 & 0.0 & \\
         & & Interpulse & 0.565 & 145.59 & \\
		 & & Precursor & 0.197 & -18.42 & \\
		\hline
	\end{tabular}
    \label{tab: table 1}
\end{table*}

The overall shift of the profile with respect to frequency indicates that residual dispersion was still present. We were able to remove the residual dispersion by fitting a $\nu^{-2}$ dispersion dependence to each individual epoch. We conservatively estimate the error on our fitting to be $0.005\,\mathrm{pc\,cm^{-3}}$ ($0.07\,\mathrm{ms}$), which corresponds to less than two pulse phase bins and is significantly smaller than the scatter-broadening time scales, so it will not affect our measurements.

As some of our observations show the presence of a delayed echo of the
intrinsic pulse profile, we expand our model by including additional
scattering screens. The pulse scatter-broadening function is expanded to allow the inclusion of an
arbitrary number $n$ of scattering screens through $s(t)=\sum_{i=0}^n
\frac{1}{\tau_i}\exp\bigl(-\frac{t-t_i}{\tau_i}\bigr)H(t-t_i)$. Here,
$\tau_i$ and $t_i$ are the pulse scatter-broadening time scale and the time offset
of screen $i$, respectively. The time offsets are referenced to that
of the first scattering screen, for which we define $t_0=0$.

\section{Results}
\label{sec:results}

\begin{figure}
	\includegraphics[width=0.49\textwidth]{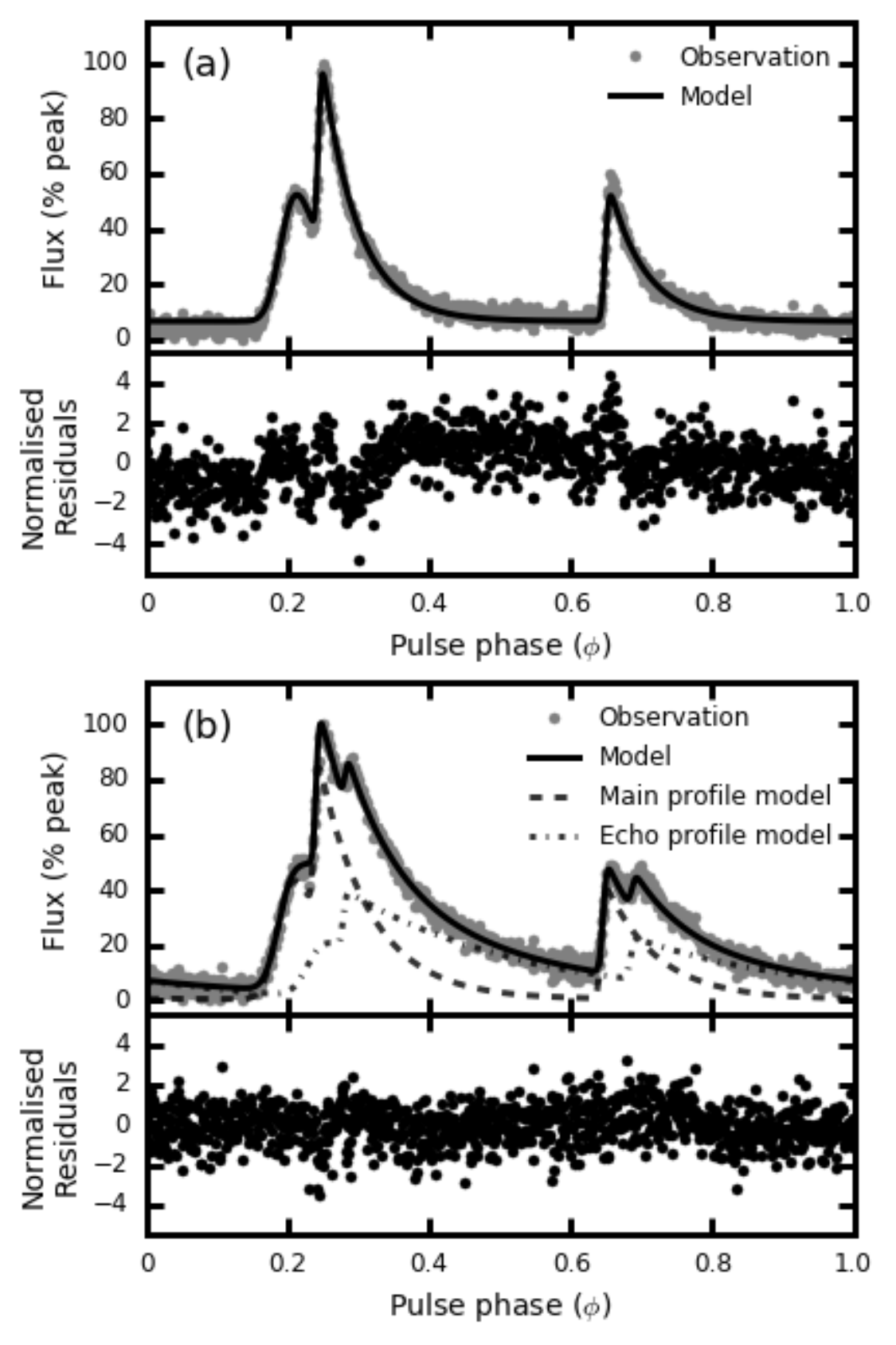}
    \caption{Examples of modelling echoes assuming (a) a single thin scattering screen and (b) two thin scattering screens. (a) and (b) show one $8\,\mathrm{MHz}$ channel of the same observations as in Fig.\,\ref{fig: figure 1}b and c respectively. The normalised residuals for the models are shown below each pulse profile, where the normalised residuals are the profile subtracted from the data and divided by the standard deviation of the off-pulse noise. The reduced $\chi^2$ values for the full fit to the pulse profile are $1.7$ and $0.9$ for (a) and (b) respectively.}
    \label{fig: figure 4}
\end{figure}

We modelled all of the observed profiles using models assuming one, and two scattering screens, some examples of models and residuals are shown in Fig.\,\ref{fig: figure 4}. For those profiles that have two clear peaks, such as in Fig.\,\ref{fig: figure 4}b, we find that a two screen model is effective and that the relative position of the second peak, or echo, to the first peak can be measured using a two screen model; however, there are only 4 observations with the echo as a clear, separate peak. As such, we cannot investigate the evolution of the separation between the echo and main profile in time.

In some cases we observe the echo to be brighter than the main profile. While the observation on 2012 December 15, see Fig.\,\ref{fig: figure 1}d, is the only profile with this feature clearly evident, there are five other profiles that suggest an echo brighter than the main profile. A possible explanation for this feature is plasma lensing, which is a refractive effect where an irregularity in a plasma screen acts as a lens. As the plasma has a negative refractive index this causes the light to diverge. Plasma lensing is observed as increases and decreases in flux density in a pattern of caustic spikes \citep{1998ApJ...496..253C} and can therefore explain a brighter echo compared to the original peaks.

\begin{figure}
	\includegraphics[width=0.49\textwidth]{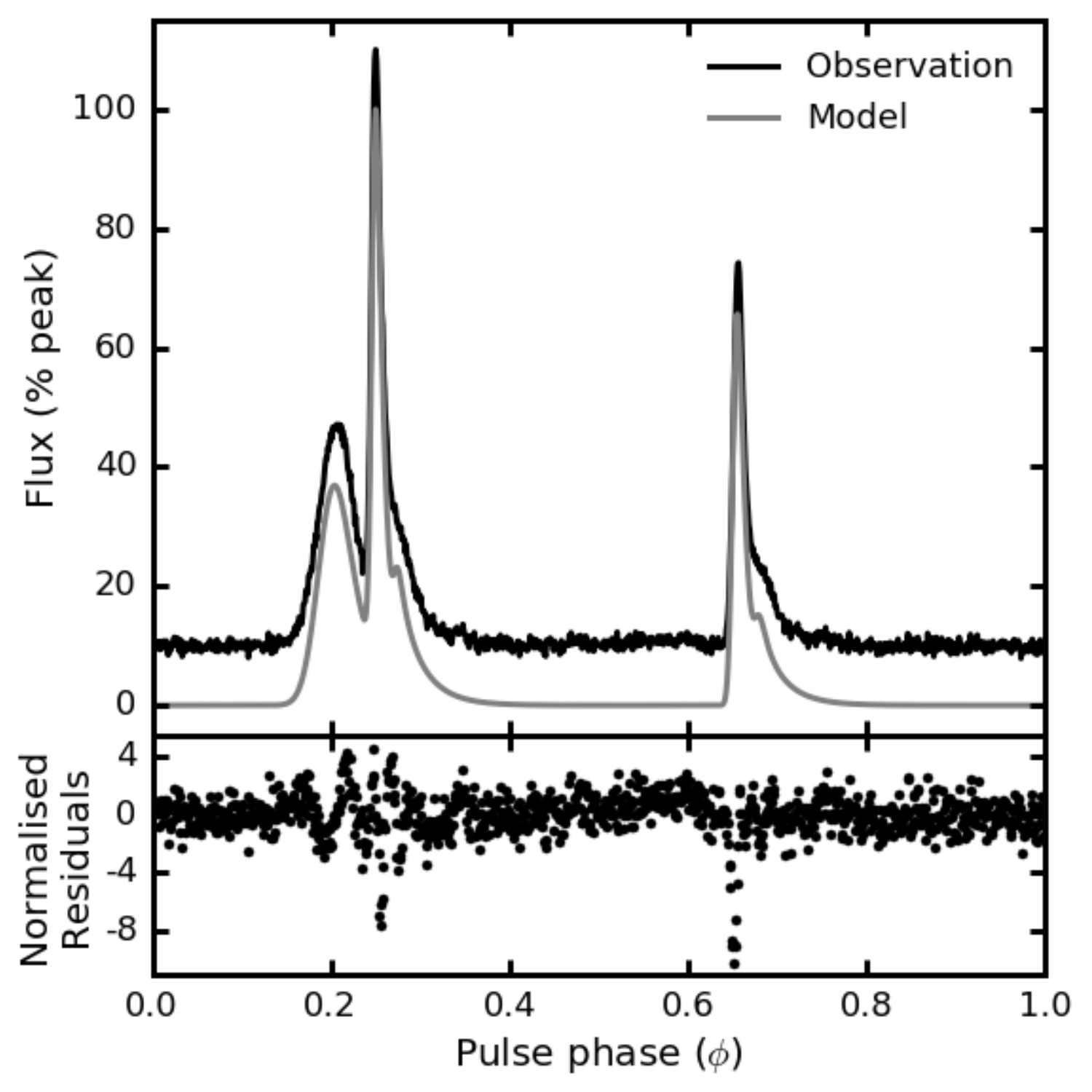}
    \caption{An example of a thin-screen approximation fit to a pulse profile with a shoulder feature. This is one $8\,\mathrm{MHz}$ channel of the same observation as shown in Fig.\,\ref{fig: figure 1}e, 2014 May 16. The model fit to the data is a two-screen model, a screen for the main profile and a screen for the shoulder. This figure shows that the smooth nature of shoulder features cannot be reproduced using the thin-screen approximation. The residuals in the lower plot show the profile subtracted from the data and divided by the standard deviation of the off-pulse noise. The residuals show that this model also does not reproduce the precursor, MP or IP correctly. The reduced $\chi^2$ value of the full fit to the pulse profile is $2.2$. The observation has been offset from the model by 10 flux units.}
    \label{fig: figure 5}
\end{figure}

Using the method described in Sec.\,\ref{sec: pulse profile modelling} we attempted to fit profiles with shoulder features using models with two thin-screens. An example of the best fit to a profile with a shoulder feature is shown in Fig.\,\ref{fig: figure 5}. In contrast to modelling exponential echoes with two thin-screens the morphology of the shoulder is not well modelled by the thin-screen approximation, in particular the model cannot reproduce the smooth nature of the shoulder.

\begin{figure}
	\includegraphics[width=0.49\textwidth]{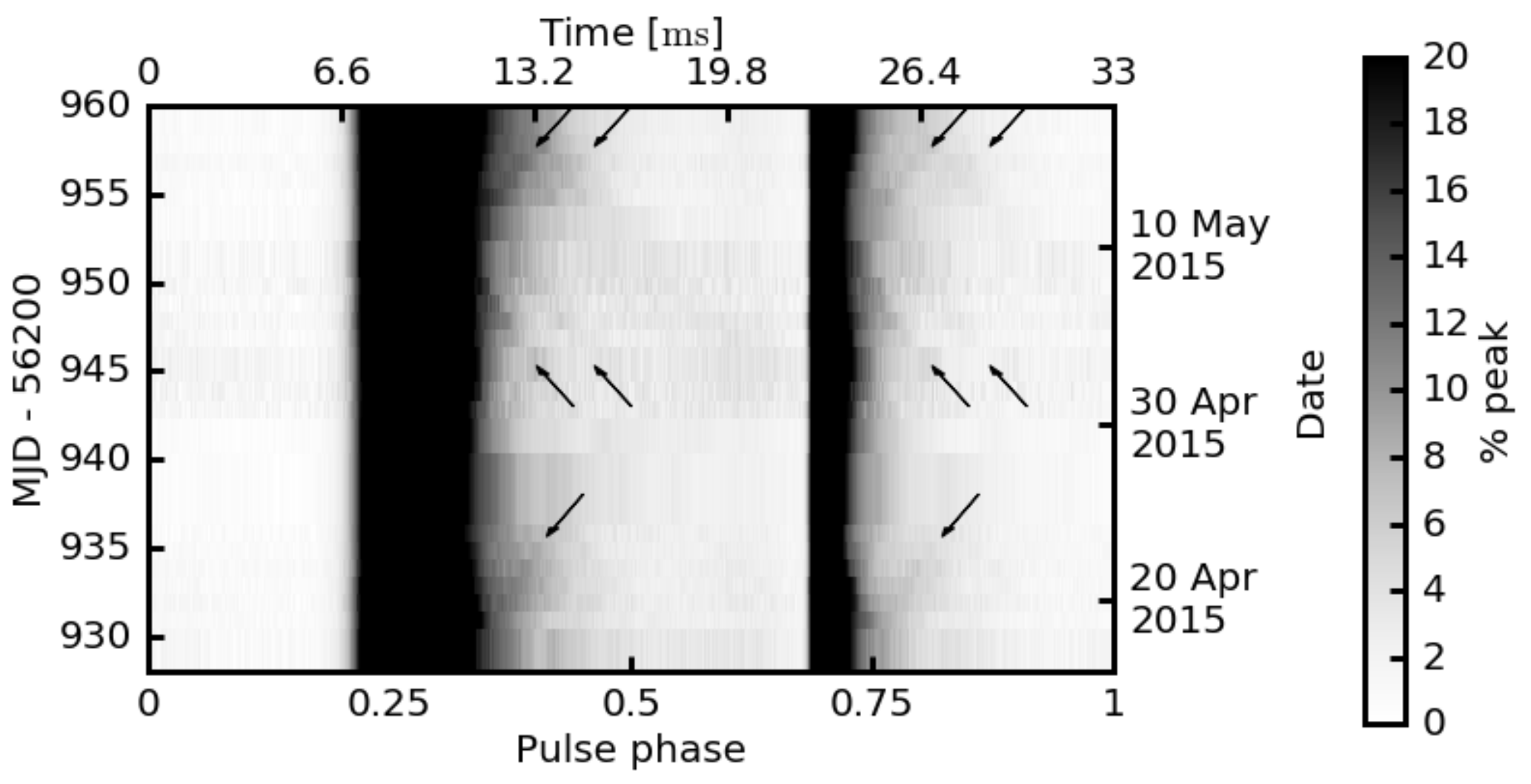}
    \caption{Observations of the Crab pulsar where bump features (i.e. non-exponential echoes) can be traced over a timescale of a couple of weeks. The bumps (indicated with arrows) approach and recede from the MP and IP over time between lines marked between 57145 and 57160. An earlier bump can be seen to recede from the MP and IP from the start of the figure for about 10 days.  The greyscale has been adjusted manually to make the bumps stand out more clearly.}
    \label{fig: figure 6}
\end{figure}

With higher cadence observations available in 2015 we are able to track bumps and shoulders in time. We find that over time the bumps approach the MP and IP, become shoulders, overlap the MP and IP, and later recede in a reverse manner; see Fig. \ref{fig: figure 6}. This indicates that shoulders and bumps are the same feature, where shoulders can be defined as bumps that are too close  in phase to the MP and IP to be separated as individual features on the pulse profile. We do not observe shoulders or bumps evolving into echoes or vice versa. A similar effect to that seen in Fig.\,\ref{fig: figure 6} was observed by \citet{2000ApJ...543..740B} and \citet{2001MNRAS.321...67L} in 1997, where the features they observed receded over 150 days after the initial approach.
\citet{2001MNRAS.321...67L} observed quadratic paths of similar features approximately eight times in daily observations of the Crab pulsar from 1984 to 1998 and therefore
suggest that they occur approximately every two years. However, although our cadence is not high enough to track the features, our data at $350\,\mathrm{MHz}$ show bump and shoulder features consistently throughout the two and a half years of our observations. This highlights that both observing frequency and cadence are important considerations for future observations of transient profile features for the Crab pulsar.

To study the pulse scatter-broadening properties we use the frequency dependent intrinsic profile and single screen approximation to determine the residual dispersion measure as well as the pulse scatter-broadening time scale and pulse scatter-broadening-frequency powerlaw ($\tau\propto\nu^{\alpha}$) dependence of all observations. Although using the single screen approximation is not optimal, it gives a first order estimate of the total extent of the pulse scatter-broadening of the pulse profile. Here, we again use those observations, which were fully a averaged in time and collapsed to 32 channels in frequency. We then model all 32 channels for each observation using single-screen models as described in Sec.\,\ref{sec: pulse profile modelling} and find the average of the resulting pulse scatter-broadening time scales, $\tau$. The DM and resulting average pulse scatter-broadening values are shown in Fig.\,\ref{fig: figure 3}, highlighting the increased pulse scatter-broadening and DM during the period where echoes are observed on the pulse profile of the Crab. The powerlaw index of all observations is $\alpha=-3.9\pm0.5$. This result is consistent with the expected powerlaw index, $-4.4<\alpha<-4$, for thin screen scattering.

\section{Discussion}
\label{sec:Discussion}

In our observations both exponential and non-exponential echoes vary in phase position (see Fig.\,\ref{fig: figure 6} and Appendix \ref{App: A}) and amplitude over time. We therefore discard the possibility that these features are intrinsic, i.e. related to other pulse profile features seen at different frequencies such as the low-frequency precursor, and instead assume that they are caused by effects in the ISM or Crab nebula.
Previous analyses of past instances of additional pulse scatter-broadening
\citep[e.g.][]{2000ApJ...543..740B,GS2011} have assumed that the additional pulse scatter-broadening and echoes are caused by a second scattering screen. The second screen is generally assumed to be from a filament, sheet, or blob in the Crab nebula and the first screen is assumed to be the ISM. Further modelling and observations are required to fully investigate the mechanisms causing exponential echoes and non-exponential echoes, and whether both features arise from the nebula, or from the ISM, and whether they are both caused by the same mechanism. To facilitate further investigation, the data is available online\footnote{\href{https://doi.org/10.5281/zenodo.1481083}{https://doi.org/10.5281/zenodo.1481083}}.
We would define the ``echoes'' observed by \citet{1975MNRAS.172...97L} as exponential echoes, and the ``echoes'' observed by \citet{2000ApJ...543..740B} as non-exponential echoes. Non-exponential echoes have been observed on a different pulsar, PSR B2217$+$47, by \citet{2018MNRAS.476.2704M} and were attributed to propagation effects the ISM.

If we assume that echoes are caused by screens or filaments in the Crab nebula we can investigate the properties of this screen, such as its size and the scattering angle.
We define the end of the screen to be the local minimum at 2013 April 22
which is indicated as region 1 in Fig.\,\ref{fig: figure 3}, with the start defined as the start of our data set.
The end of this region was chosen
because the pulse scatter-broadening reaches a minimum
and we see no obvious echo feature in this observation. Choosing these limits means that the 
screen was crossing the line of sight to the pulsar for at least
$\approx139\,\mathrm{days}$.

Given our observing boundaries, and if we assume that the velocity of the filament relative to the Crab pulsar is due completely to the proper motion of the pulsar \citep{GS2011},
$120\,\mathrm{km}\,\mathrm{s}^{-1}$ 
\citep{2008ApJ...677.1201K}, this filament is found to extend for at 
least $9.6\,\mathrm{AU}$.
The period of additional pulse scatter-broadening observed by \citet{GS2011} had a shorter duration than the pulse scatter-broadening period discussed above. They estimated the filament causing the pulse scatter-broadening to extend for 
$\approx10\,\mathrm{days}$ corresponding to $0.72\,\mathrm{AU}$.

\begin{figure}
	\includegraphics[width=0.49\textwidth]{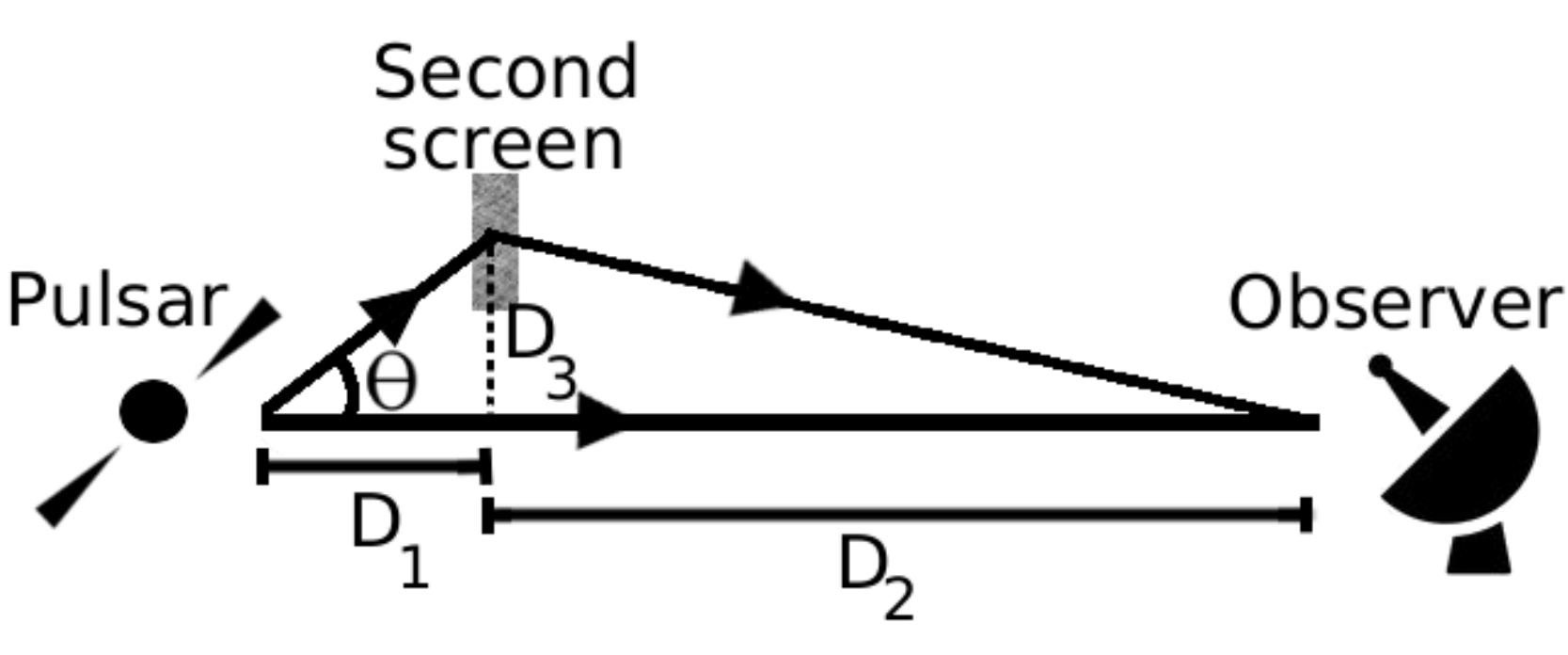}
    \caption{Diagram showing scattering angle and distance between the pulsar and the
    		 observer.}
    \label{fig: figure 7}
\end{figure}

If we assume that the second screen is within the Crab nebula then
we can calculate the scattering angle.
The scattering angle, $\theta$ in Fig.\,\ref{fig: figure 7}, is
the angle between the direct line of sight to the pulsar ($D_1+D_2$) and the direction to the screen producing the echo. If distance $D_2\gg D_1$, as is 
the case if the second screen is in the nebula, then we can assume that $D_1+D_2\approx D_2$. This means that $\theta$ can then be found using:
\begin{equation}
    \Delta t \approx \frac{D_{1} \theta^{2}}{2c}
    \label{eq:theta equation}
\end{equation}
where $\Delta t$ is the time delay between the arrival of the main profile and the echo,
$D_1$ is the distance shown in Fig.\,\ref{fig: figure 7}, and $c$ is the speed of light.

At $350\,\mathrm{MHz}$ on 2012 December 18 the shift in
time between the main profile and the echo is $\Delta t\,\approx\,1.5\,\mathrm{ms}$. 
The Crab pulsar is $2\,\mathrm{kpc}$ from the Earth \citep{2008ApJ...677.1201K} and
the nebula is an ellipsoid of $\approx 8.3\times10^5\,\mathrm{AU}\,\times\,6.2\times10^5\,\mathrm{AU}$
\citep[][]{2014BASI...42...47G}. The termination shock radius
of the cold pulsar wind is at $\approx2.1\times10^4\,\mathrm{AU}$ from the pulsar \citep{1975MNRAS.172...97L}. If we assume the screen is somewhere between the pulsar wind termination shock
and the outer edge of the nebula we find that $\theta$ is between $3.52\,\mathrm{arcsec}$
and $0.56\,\mathrm{arcsec}$. This means that $D_3$ would be between 
$2.2\,\mathrm{AU}$ and $0.35\,\mathrm{AU}$.

While we were able to use two-screen models using the thin-screen approximation to model observations with exponential echoes, we were unable to model non-exponential echoes using a thin screen model because they have a smooth Gaussian-like structure compared to exponential echoes. It is possible that the thin-screen assumption does not hold for these features, but further investigation and modelling are required to understand these features.

In our observations and the 1974 observations it is not clear whether exponential echoes evolve in time by first approaching the main profile, overlapping the main profile, then later receding, in a similar way to non-exponential echoes. We expect the evolution in time to be similar, assuming that both exponential and non-exponential echoes are caused by material crossing the line of sight to the pulsar. Higher cadence observations are required to investigate this, but due to the increased scattering when exponential echoes occur it may be difficult to track how they vary over time.

\section{Conclusions}
\label{sec:Conclusion}

We observe pulse scatter-broadening, exponential echoes, and non-exponential echoes consistently in WSRT observations at $350\,\mathrm{MHz}$ of the Crab pulsar over two and a half years. We find that we can model pulse scatter-broadening and exponential echoes with a thin-screen scattering approximation; however, this model cannot replicate non-exponential echoes. We observe that shoulders and bumps are the same non-exponential echoes, but at different phases relative to the main profile. Non-exponential echoes approach the MP and IP over time and later recede, similar to previous observations. As this is only the third time echoes have been observed on Crab pulse profiles, further investigation is required to determine whether exponential echoes and non-exponential echoes are produced by the same inhomogeneities in the ISM or Crab nebula and whether they are caused by the same scattering mechanism. We rule out intrinsic profile changes as exponential and non-exponential echoes vary in phase position and amplitude over time.

Our observations show that daily monitoring of the Crab pulsar at $\sim350\,\mathrm{MHz}$ would be a useful tool for further investigating transient features caused by propagation effects. It would be interesting to investigate these features using Crab pulsar giant pulses as giant pulses are intrinsically very bright and narrow. They also have only a single component, which would make the effects of pulse scatter-broadening more apparent and would make fitting easier, even when the pulse scatter-broadening time is much greater than the pulse period. Both higher cadence observations at $\sim350\,\mathrm{MHz}$ and investigations using giant pulses would lead to further insight on which material in the Crab nebula causes these effects, how often they occur, and their evolution in time.

\section*{Acknowledgements}

The Westerbork Synthesis Radio Telescope is operated by the
Netherlands Institute for Radio Astronomy (ASTRON) with support from
The Netherlands Foundation for Scientific Research (NWO). 
LND and BWS acknowledge support from the European Research Council (ERC) under the
European Union's Horizon 2020 research and innovation programme (grant agreement No
694745).
We would like to thank Andrew Lyne, Marten van Kerkwijk, and James McKee for useful and interesting discussions. LND would like to thank the ASTRON/JIVE summer studentship programme for a fellowship in 2014.
We thank the referee, assistant editor, and scientific editor for their constructive comments on the manuscript.

%%%%%%%%%%%%%%%%%%%%%%%%%%%%%%%%%%%%%%%%%%%%%%%%%%

%%%%%%%%%%%%%%%%%%%% REFERENCES %%%%%%%%%%%%%%%%%%

\nocite{Hunter:2007}
\nocite{scipy}
\nocite{2004PASA...21..302H}
\nocite{1977ApJ...214..214I}
\bibliographystyle{mnras}
\bibliography{CrabBib}

% %%%%%%%%%%%%%%%%%%%%%%%%%%%%%%%%%%%%%%%%%%%%%%%%%%

% %%%%%%%%%%%%%%%%% APPENDICES %%%%%%%%%%%%%%%%%%%%%

\appendix

\section{Crab pulse profiles}
\label{App: A}

Here we include the pulse profiles of the 84 WSRT observations of the Crab pulsar. All profiles are WSRT $350\,\mathrm{MHz}$ observations of the Crab and have been averaged in time and collapsed in frequency. Every pulse profile has been rotated such that the maximum peak is at $\phi=0.25$, and scaled such that the peak value is 1.0.

\begin{figure*}
	\includegraphics[width=0.9\textwidth]{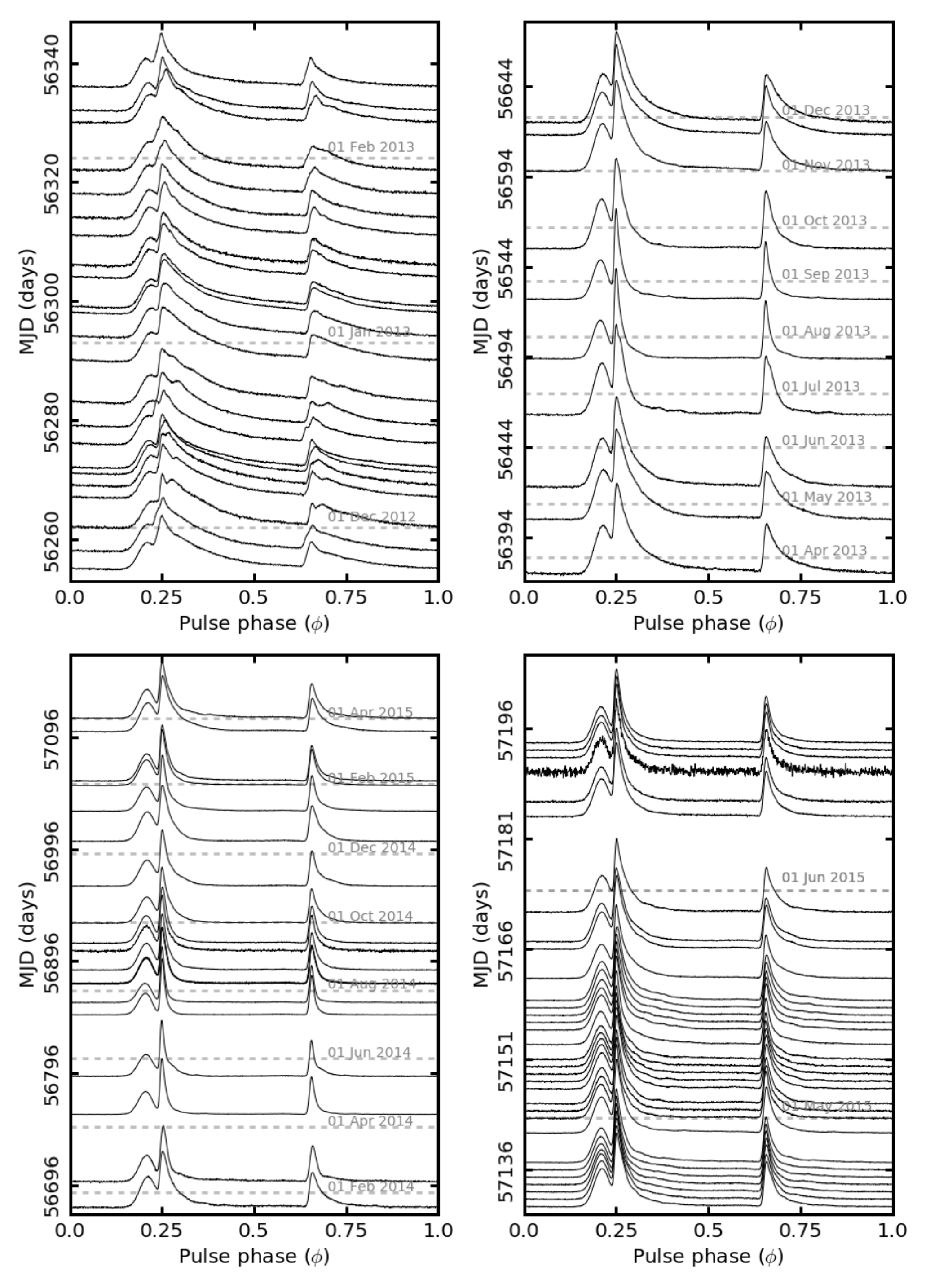}
     \caption{Pulse profiles of the Crab pulsar at $350\,\mathrm{MHz}$ observed by WSRT from late 2012 until mid 2015. Each line represents one observation, and each profile has been scaled such that the maximum is 1.0 and the minimum is 0.0, and rotated such that the main peak is at $\phi=0.25$.}
     \label{fig:stacked profiles 1}
\end{figure*}

%%%%%%%%%%%%%%%%%%%%%%%%%%%%%%%%%%%%%%%%%%%%%%%%%%

% Don't change these lines
\bsp	% typesetting comment
\label{lastpage}
\end{document}